\documentclass[aps,pre,groupedaddress,onecolumn]{revtex4}
\pdfoutput=1

\newcommand{\bbar}{b{\hspace{-6pt}^{^{\_\hspace{-3pt}\_}}}}

\usepackage{hyperref}
\usepackage{graphicx}
\usepackage{natbib}
\usepackage{doi}

\begin{document}
%bibliographystyle{plainnat}
%\bibliography{myrefs}
%%\bibliographystyle{abbrvnat}
%%\bibliographystyle{unsrtnat}
\expandafter\ifx\csname urlprefix\endcsname\relax\def\urlprefix{URL }\fi

\DeclareGraphicsExtensions{.pdf, .jpg}

\title{\large
 Quantum consciousness in warm, wet, and noisy brain
    }

\large
\author{Valeriy I. Sbitnev}
\email{valery.sbitnev@gmail.com}

\affiliation{B.P.Konstantinov St.-Petersburg Nuclear Physics Institute, NRC Kurchatov Institute,
     Gatchina, Leningrad district, 188350, Russia;\\
      Department of Electrical Engineering and Computer Sciences, University of California at Berkeley, Berkeley, CA 94720, USA
      }

% Comments: 20 pages, 11 figures
% Subjects: Quantum Physics (quant-ph)
% Subjects: Fluid Dynamics (physics.flu-dyn)
% Subjects: Neurons and Cognition (q-bio.NC)

\date{\today}

\begin{abstract}
The emergence of quantum consciousness stems from dynamic flows of hydrogen ions in brain liquid. This liquid contains vast areas of the fourth phase of water with hexagonal packing of its molecules, the so-called exclusion zone (EZ) of water.  The hydrogen ion motion on such hexagonal lattices shows as the hopping of the ions forward and the holes (vacant places) backward, caused by the Grotthuss mechanism. By supporting this motion using external infrasound sources, one may achieve the appearance of the superfluid state of the EZ water.
Flows of the hydrogen ions are described by the modified Navier-Stokes equation. It, along with the continuity equation, yields the nonlinear Schr\"odinger equation, which describes the quantum effects of these flows, such as the tunneling at long distances or the interference on gap junctions.
 \\

{\bf Keywords:} 
  brain liquid;  hydrogen ion; Grotthuss mechanism; quantum flow; exclusion zone; superfluidity;
  Navier-Stokes; Schr\"odinger equation; Bohmian mechanics;
  gap junction; interference

%\\ PACS numbers: {03.65.Yz, 47.10.ad, 87.85.dm}

\end{abstract}

\maketitle

%\Large

\section{\label{sec1}Introduction}

How consciousness arises from physical or material activity in the brain is the hard problem 
in the study of consciousness~\citep{Nader2015, Tarlaci2010}. 
 This remains a big question~\citep{Tarlaci2013, Jahn1981, Prigogine1994, Beck1994, StappEtAl2005, Georgiev2012, Georgiev2015, Tegmark2015},
and we find a wide spectrum of philosophical opinions ranging from idealism, materialism, Eastern mysticism, to religious sacred texts. 
In any case, we face the dualism problem, which deals with the relationship of the mind and the body - the mind-body problem. 
This problem has fascinated the humanity since ancient times.

  The clearest understanding of the mind-body problem was formulated 
  by Ren\'e Descartes~\citep{Skirry1995},
   a seventeenth-century philosopher and scientist.
  He provided the best-known version of dualism.
  Descartes was the first to bind the mind with consciousness and self-awareness and distinguished this from the brain, 
which houses intelligence.
The main assumption was that consciousness, like a non-material ephemeral cloud, is outside the brain and is detected by a special organ called the pineal gland~\citep{PinealGland2005} 
(also called epiphysis cerebri or the "third eye").  
It is located near the center of the brain between the two hemispheres. 
This reddish-gray pineal body is the size of a pea (8 mm in humans). 
 It often contains calcifications (brain sand) which are easily identified on X-ray images of the brain.
 
  Let us take a look at Descartes' doctrine from the viewpoint of scientists of the 21st century. 
First, we may think of the non-material ephemeral cloud as a wave function which carries knowledge of phenomena occurring in the world. 
Such a view is consistent with the Copenhagen interpretation of quantum mechanics~\citep{Copenhagen2014}
 according to which the wave function provides knowledge of phenomena, but does not point to 'really existing objects'. 
Manifestation of knowledge in the brain, as was noted above, is achieved due to the pineal gland, "third eye". 
This gland is a detector in our understanding where the wave function collapses as a result of measurement. 
 
 A radical solution of such a quantum measurement as a process of consciousness was proposed by Stuart Hameroff, together with Roger Penrose. Their model (orchestrated objective reduction: Orch OR~\citep{Hameroff1998, Hameroff2014, PenroseHameroff2011, HameroffPenrose2014a})
  suggests that quantum superposition and a form of quantum computation occur in microtubules  -- cylindrical protein lattices of the cell cytoskeleton within the brain's neurons. 
 Here the microtubules play a role of detectors on which a collapse of wave functions occurs. The latter describes the effects of quantum gravity on the Planck-scales of spacetime geometry. 
 Their opinion is that there is a connection between the basic structure of the Universe and the brain's biomolecular processes.
 
 As soon as the 'Orch OR' model was published, it was criticized by~\citet{Tegmark2000}, 
 whose primary remarks concerned the following egregious discrepancies: (a)~the collapse of the wave function is  much shorter that of relevant dynamic timescales of neuron firings; and (b)~wet warm brain, working at room temperature, cannot provide supporting quantum computations. Thermal noise of the brain completely drowns out such delicate computations.
 
 Analogous remarks have recently been stated by a number of scientists.
 It is stated that the brain is a warm, wet, and noisy organ as say~\citet{JumperScholes2014}
  and therefore  'Orch OR' model for human consciousness is not scientifically justified~\citep{ReimersEtAl2014}. 
\citet{Tuszynski2014} 
emphasizes that the gravitational interactions are many orders of magnitude weaker than
 even those of thermal noise in neurons. For that reason, the 'Orch OR' theory has a fatal flaw. He remarks, nevertheless, that one can foresee progress  in  bridging  the  gap  between  nanoscience  and  consciousness where the  microtubules can serve as storage of memory.

Taking into account these criticisms, let's look at the Hameroff-Penrose theory from the other side, the side mentioned by Tuszynski. 
By comparing the Hameroff-Penrose solution with  Descartes' idea, one can see that instead of the pineal gland having 8 mm in diameter, an enormous field of the  microtubules occupying the whole brain  and wetted everywhere by the cerebral liquid organizes the process of consciousness. 
So we can state a hypothesis, that the warm, wet, and noisy brain, containing about 90\% water, also contains microtubules 
occupying the nervous cells everywhere densely~\citep{ Hameroff2014, HameroffPenrose2014a}.
They are not only 'scaffolding' of cells but can also serve as warehouse memory and memristors~\citep{Chus1971, Chua2011}, 
downloaded by heavy ions, such as calcium ions.

The article is organized in the following manner. 
Sec.~\ref{sec2} gives a general picture ranging from the dark energy and vacuum fluctuations up to a place of brain in this world picture, water basin of the brain, and transport of the hydrogen ions in water.  
Sec.~\ref{sec3}  introduces the modified Navier-Stokes equation describing flows of  hydrogen ions in a slightly saltish water that primary is close to the cerebrospinal liquid. An aim in Sec.~\ref{sec4} is getting the Schr\"odinger equation from the above Navier-Stokes equation.
A separate subsec.~\ref{sec4A} is devoted to the Grotthuss mechanism describing motion of the hydrogen ions on the hexagonal lattice of packed water molecules, fourth state of water,  superfluid state of EZ water.
 Sec.~\ref{sec5} introduces the Feynman  path integral technique for solving the Schr\"odinger equation
 and describes emergence of interference patterns.
 Concluding Sec,~\ref{sec6} discusses consciousness as a physical phenomenon emerging in an attempt to achieve a goal. It is a special form of movement in the brain memory space.

\section{\label{sec2}Amazing coincidences}

Living brain is a biological organ which operates in a slightly salty liquid environment at room temperature. 
The human brain is made up of 95\% water.
Electrochemical activity of the brain is taking place in a wet and warm environment. 
Consequently, the quantum fluctuations are sinking against a background of thermal fluctuations of biochemical components of the brain. 
It can mean that the magnitude of the Planck constant, $h$, may be much less than $k_{_{\rm B}}T{\delta\tau}$, where $T$ = 298 K is the room temperature and ${\delta\tau}$ is the characteristic relaxation time of the biochemical component in question. 
For that reason,  we chose the thermal action parameter, $k_{_{\rm B}}T{\delta\tau}$, instead the Planck constant, which is assembled from the Boltzmann constant $k_{_{\rm B}}$, the room temperature,
 $T$ = 298 K, and the characteristic relaxation time ${\delta\tau}$ for some kind of  basic chemical substance.
 
 Most widespread chemical substance in the living body is liquid water~\citep{Chaplin2016}. 
In this fluid medium, the hydronium ion, H$_3$O$^+$, is a carrier of protons. 
The average lifetime of the hydronium ion in water is about $2\cdot10^{-13}$ s~\citep{Bell1973}.
Hydrogen ion, that is proton, is considered here as a bit of information transmitting across the cerebral liquid of brain by so-called the Grotthuss mechanism~\citep{Chaplin2016b}
 (this mechanism will be considered later). 
It accomplishes Brownian motion which is adequately described by the Wiener process~\citep{Nelson1967, Nelson1985}. 
In turn, motion of ensemble of protons can be described by means of the Navier-Stokes equation
together with the continuity equation.

Proton exceeds the electron mass on about 2000 times. 
It means that proton is a more inertial particle than electron, and, consequently, more robust for thermal fluctuations.
For that reason, we adopted the hydrogen ion as the unit of thermal motion.
In water solutions, the hydrogen ion resides in a bound state as hydronium, H$_3$O$^+$, 
with the average lifetime of about $10^{-13}$~s~\citep{Bell1973}. 
So motion of the hydrogen ion occurs accidentally by means of jumps in the averaged time of about 
${\delta\tau} = 2\cdot10^{-13}$~s. We choose this value and proclaim that the following thermal action parameter 
\begin{equation}
 b = k_{_{\rm B}}T{\delta\tau}
\label{eq=1}
\end{equation}
imitates the Planck constant $h$.
 This parameter is not constant but depends on temperature of the tank, composition of the solution in the tank, and so forth. 
 In the case of $T$ = 298 K and ${\delta\tau} = 2\cdot10^{-13}$~s this parameter has a value $8.238\cdot10^{-34}$~J$\cdot$s, 
 whereas the Planck constant, for comparison, is $6.626\cdot10^{-34}$~J$\cdot$s. 
 As seen, these values  are almost equal. 
It should be noted, that in the case of the human body temperature, $T\approx310$ K, the thermal action parameter and  the Planck constant would be equal each other, if the lifetime were close to $1.55\cdot10^{-13}$~s. 

   Although the Planck constant and the  thermal action parameter are almost equal, these parameters have qualitative differences.  The Planck constant arises as a character representing a quantum of radiation of absolutely black body, whereas the   thermal action parameter represents an action of thermal migration of a single proton from one hydronium ion to the other. 
 
 In the fluid medium, as a basic speed of matter transfer we choose the speed of sound proposed by~\citet{BradyAnderson2014}. 
 In a saline water solution at room temperature, $T = 298$~K, the speed is equal to $c_{s}=1508$~m/s. 
 The speed is taken from the formula of~\citet{Wilson1960}.
% (see  {URL \url{http://www.akin.ru/spravka_eng/s_i_svel_e.htm}}).
    Observe that
\begin{equation}
    m_{*} = {{k_{_{B}}T}\over{c_{s}^2}} \approx 1.81\cdot10^{-27}~~{\rm kg}
\label{eq=2}
\end{equation}   
is  slightly larger then the proton bare mass $m = 1.6726\cdot10^{-27}$~kg.
 The mass increase is due to a resistance experienced by the hydrogen ion when it moves into the liquid solution.
Specifically, in the case of human body temperature, $T=309.75$ K, sound speed is $1535$ m/s, and the mass $m_{*}$ is also close to $1.81\cdot10^{-27}$ kg. However, we shall deal with water near room temperature, since only at this temperature will we have a full set of essential parameters for our evaluations. They are gathered in Table 1.
\begin{figure}[htb!]
  \centering
  \begin{picture}(200,135)(-10,25)
      \includegraphics[scale=0.30, angle=90]{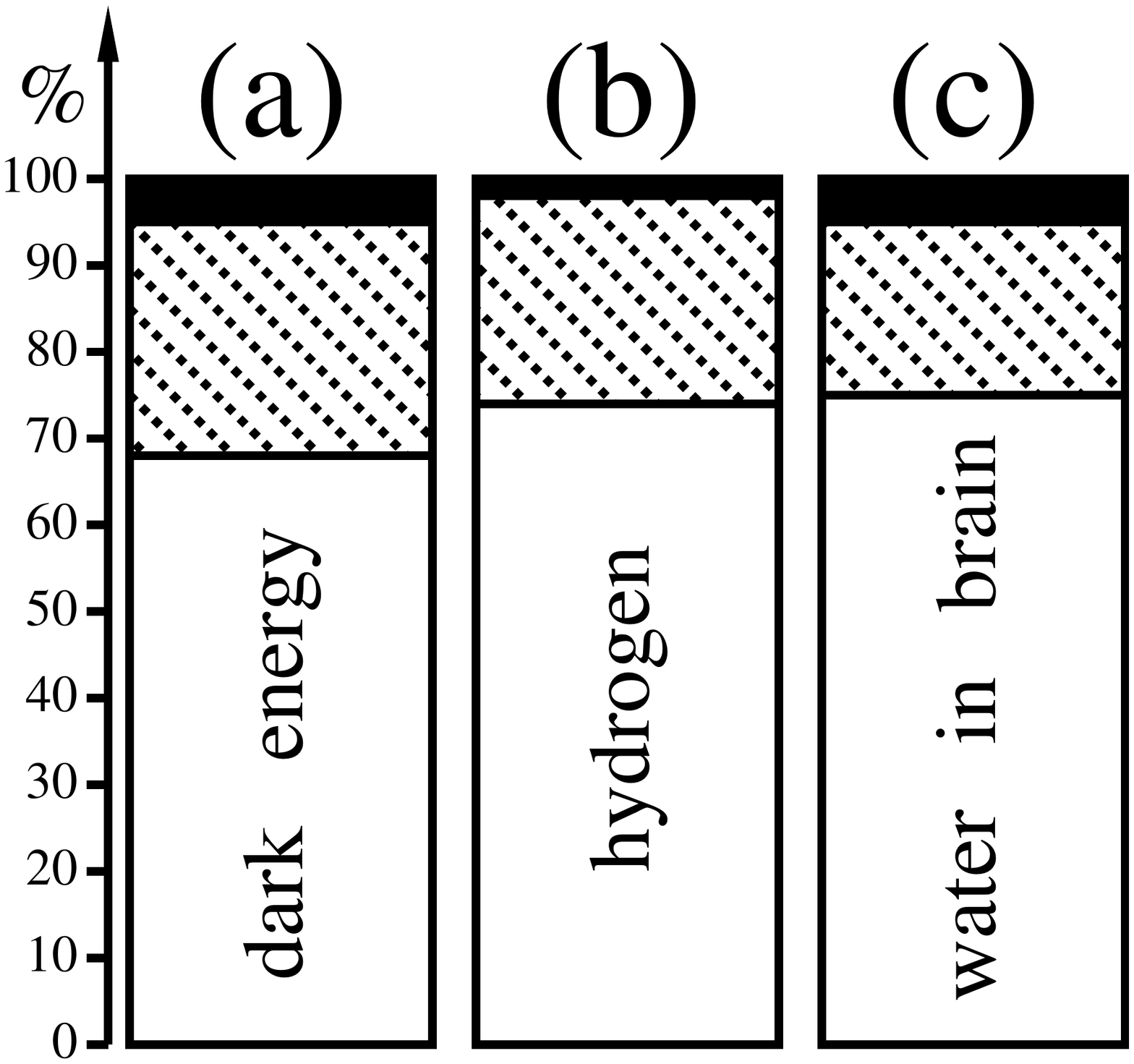}
  \end{picture}
  \caption{
Percent content of different substances in Nature: (a)~Universe consists of  dark energy, 68\% (white rectangle,~WR),  dark matter, 27\% (hatched by beveled dotted rectangle,~DR),  baryonic matter, 5\% (black rectangle, BR); (b)~baryonic matter contains hydrogen, 74\% (WR),  helium, 24\% (DR),  other heavy atoms, 2\% (BR); (c)~brain contains water, 75\% (WR),  lipids and proteins, 20\% (DR),  soluble organic matter, inorganic salts, carbohydrate 5\% (BR).
  }
  \label{fig=1}
\end{figure}

  It should be noted that
  hydrogen is the most abundant chemical element in the Universe (74\% and about 24\% of all baryonic matter compose hydrogen and helium, respectively). Curiously, all baryonic matter 
   composes only about 4\% against all dark matter and energy, according to scientists, Fig.~\ref{fig=1} . 
  One may say, that a boundless ocean of the dark substance that occupies about 95\% of total mass-energy content spans us, spreads across us, and stretches everywhere.  

To summarize, we can represent the content of substances in Nature in the percentages shown in Fig.~\ref{fig=1} . 
Bar~(a) shows that only 4\% of baryonic matter exists in our Universe, as opposed to background of the dark energy and dark matter which are about 96\%~\citep{Sbitnev2016a}. The baryon matter contains mostly hydrogen and helium, about 98\%, see bar~(b). Hydrogen, together with oxygen, forms a water which is a mother of life on Earth. Amazingly, the mammalian brain contains about 75\% of water~\citep{Tarlaci2013}, see bar~(c). 

The water is a main liquid medium in the brain, where all important events occur.
\begin{figure}[htb!]
  \centering
  \begin{picture}(200,230)(20,5)
      \includegraphics[scale=1.2]{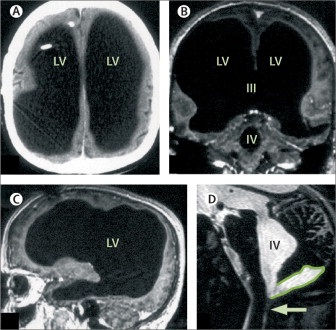}
  \end{picture}
  \caption{
Massive ventricular enlargement, in a patient with normal social functioning: 
(A), (B), (C) MRI with gadolinium contrast at different cross-sections; (D) T2-weighted MRI. LV=lateral ventricle. III=third ventricle. IV=fourth ventricle. Arrow points to Magendie foramen. The posterior fossa cyst is outlined in (D). The figure is taken from~\citep{FeuilletEtAl2007}.
  }
  \label{fig=2}
\end{figure}
Although dendrites and axon terminals of neurons of the brain penetrate through all brain space densely, there are spaces relatively free of the nervous filaments. These spaces are ventricles of the brain filled by the cerebral liquid. In medical practice there is a case where a 44-year-old patient with postnatal hydrocephalus of unknown cause~\citep{FeuilletEtAl2007}
 showed via magnetic resonance imaging (MRI) that his brain had hypertrophied brain ventricles,~Fig.~\ref{fig=2}. The deficit of the filamentous organization demonstrates massive enlargement of the lateral, third, and fourth ventricles, with a very thin cortical mantle and a posterior fossa cyst. Surprisingly, however, this patient possesses OK normal social functions, and has an intelligence quotient (IQ) of about~75. This example provides an indirect hint of the fact that the cerebral liquid, a slightly brackish water, has a direct relationship to cognitive functions of the brain.

At room temperature, the liquid water is represented as consisting of many fluctuating hydrogen-bonded clusters~\citep{Chaplin2016a}. 
The hydrogen bond is strong enough to maintain the coupling of atoms during some time under thermal fluctuations.
The following working hypothesis concerning the hydrogen bonding of liquid water is adopted here: water consists primarily of a mixture of clusters of water molecules with different degrees of hydrogen bonding in an equilibrium.
Under thermal fluctuations some hydrogen couplings are broken but other arise. On average, the equilibrium distribution of different cluster sizes is maintained. 
\begin{figure}[htb!]
  \centering
  \begin{picture}(180,70)(10,20)
      \includegraphics[scale=0.5]{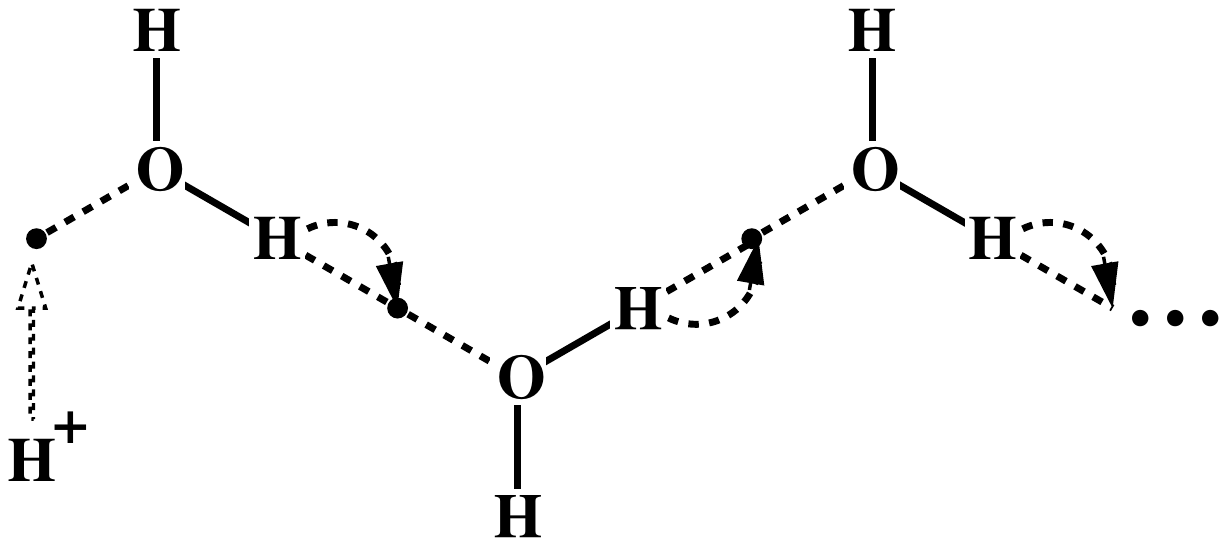}
  \end{picture}
  \caption{
Diagram illustrating the hydrogen-bonded chain mechanism for proton migration (Grotthuss mechanism)~\citep{Decoursey2003}:
a~proton enters the chain from the left side and then, as a result of the series of proton hops indicated by the arrows, a proton exits the chain on the right side. This chain represents a hydrogen-bonded 'water wire'~\citep{Chaplin2016a}.
  }
  \label{fig=3}
\end{figure}
Fig.~\ref{fig=3} illustrates the hydrogen-bonded chain mechanism~\citep{Decoursey2003},
 called the Grotthuss mechanism, by means of which protons tunnel from one water molecule to the next  via hydrogen bonding~\citep{HassanaliEtAl2013, Chaplin2016b}.
  Surprisingly,  excess protons can create their own pathways, `water wires', before protons can migrate along~\citep{PengEtAl2015}.
\begin{figure}[htb!]
  \centering
  \begin{picture}(180,245)(10,30)
      \includegraphics[scale=0.75]{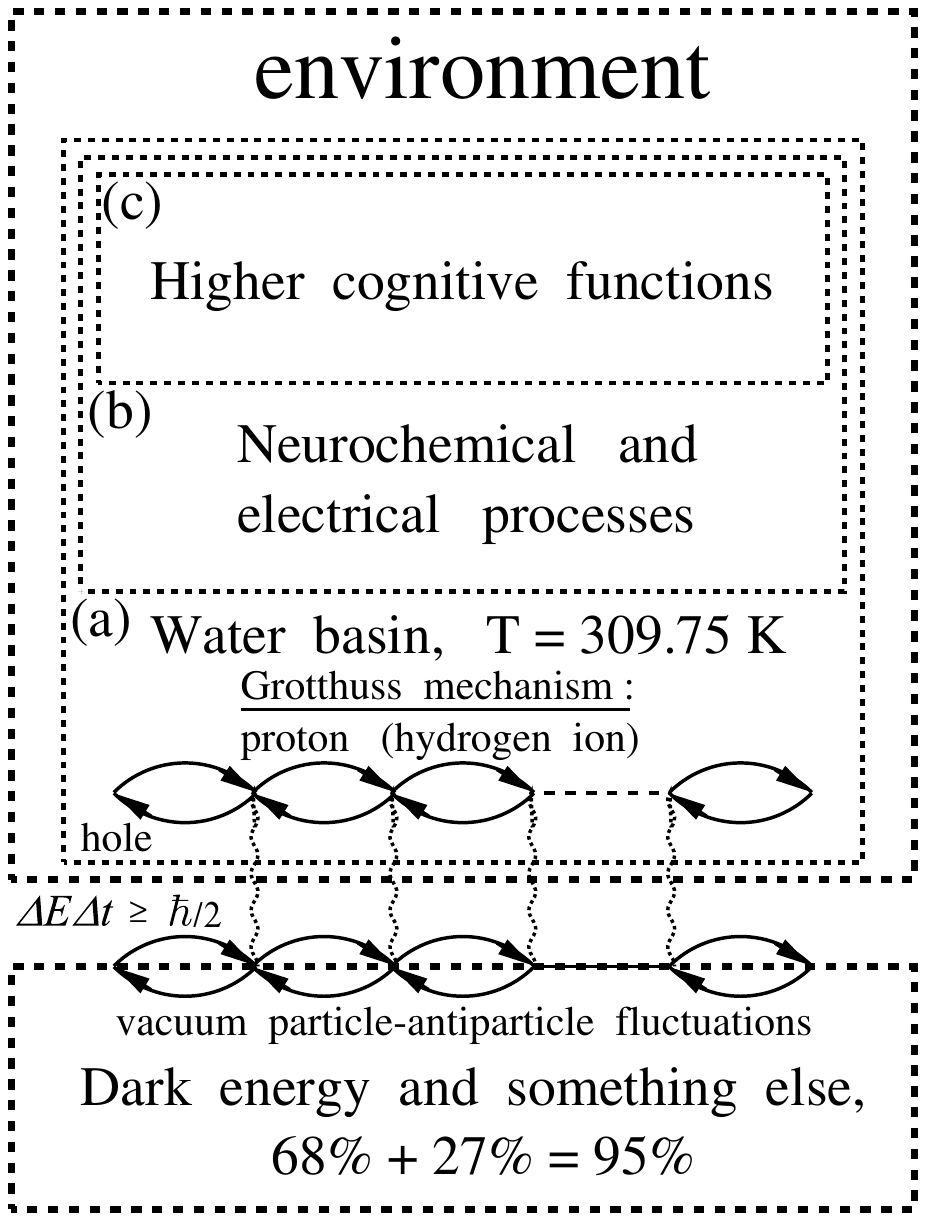}
  \end{picture}
  \caption{
A three-level diagram of brain organization shows: (a) a water basin containing all the other levels; (b) nervous tissue and glial cells, supporting electrical activity and neurochemical background; and (c) higher cognitive functions creating consciousness as a paradox~\citep{Allakhverdov2000b}. All of this is backed by a dark substance - 
the "superfluid quantum space"~\citep{Sbitnev2016a, Sbitnev2016b, Fedi2016}.
  }
  \label{fig=4}
\end{figure}

To emphasize the place of proton in brain consciousness functions, we show a rough diagram of brain organization, shown in Fig.~\ref{fig=4}. A conditional diagram of brain organization consists of three levels~\citep{Tarlaci2010}: (a) a `water basin' containing all the other levels; (b) a level consisting of many neurons and glial cells, which organize the electrical activity of the nervous cells against a background of the neurochemical medium prepared by special nuclei and cells; and (c) a level expressing higher cognitive functions, which provide adequate behavior of a living organism in the social environment~\citep{Allakhverdov2000b}.

It is noticed earlier, that the thermal action parameter (\ref{eq=1}) and the Planck constant are almost equal each other. It means, that exchange of the hydrogen ion energy with the vacuum zero-point energy can be available as well. It should be noted  in this regard, that the quantum mechanical zero-point energy is mentioned also by Beck and Eccles, in article~\citep{BeckEccles1992} entitled "quantum aspects of brain activity and the role of consciousness". The mass of the quasi-particle, which they adopt in this article, is the mass of the hydrogen atom.

The vacuum zero-point energy fluctuations happen on surface of  a vast ocean of energy called by the dark energy, Fig.~\ref{fig=4}, or more specifically the "superfluid quantum space", the name proposed by~\citet{Fedi2016}. It contains virtual particle-antiparticle pairs, which are created and annihilated again and again by staying in a continuous vortex dance. The pairs possess integer spin and therefore form the Bose-Einstein condensate covering the entire Universe~\citep{AlbaretiEtAl2014b, DasBhaduri2015}.
Thus, the hydrogen ion (proton) has the possibility of tunneling through the space on  huge distances.

Let us look at the proton migration mechanism from the perspective of the de Broglie-Bohm theory~\citep{deBroglie1987, Bohm1952a, Bohm1952b}.
 In this case, the `water wires' look like the Bohmian trajectories forming every time the need arises. But first we shall show the hydrodynamical equations describing motion of hydrogen ions in the water medium.
Factually, the balance equation~\citep{DudkinSbitnev1998} is written down initially,
that  leads further to the Naveir-Stokes equation + the continuity equation.

 \section{\label{sec3}Description of motion of a fluid with the hydrogen ions} 
 
 A stream of hydrogen ions in a water solution under forces acting on it can be described by the Navier-Stokes equation. We begin  with the modified Navier-Stokes equation~\citep{Sbitnev2016b}
\begin{equation}
 \rho_{_{M}}\biggl(
 {{\partial {\vec {\mathit v}}}\over{\partial\,t}}
 + ({\vec {\mathit v}}\cdot\nabla){\vec {\mathit v}}
       \biggr) 
  =      {\vec f({\vec r},t)}    
   \;-\;
   \underbrace{ \rho_{_{M}}\nabla (P/\rho_{_{M}})
 \; +\; \mu\,\nabla^{\,2}{\vec {\mathit v}} }.
\label{eq=3}
\end{equation} 
For completeness, we add the continuity equation
\begin{equation}
 {{\partial\,\rho_{_{M}}}\over{\partial\,t}} +(\nabla\cdot{\vec{\mathit v}})\rho_{_{M}} = 0.
\label{eq=4}
\end{equation}
 In these equations  ${\vec{\mathit v}}$ is the flow velocity.
 
   The mass density $\rho_{_{M}}$ is defined as
\begin{equation}
 \rho_{_{M}} = {{M_{}}\over{\Delta V}}={{m_{*}N}\over{\Delta V}} = m_{*}\rho.
\label{eq=5}
\end{equation}
Here, the mass $M_{}$ of the fluid filling the volume $\Delta V$ is equal to
the product of the number,  $N$, of identical sub-particles with the mass $m_{*}$ populating this volume~\cite{JackiwEtAl2004},
and $\rho=N/{\Delta V}$ is the probability density of these sub-particles in this volume.
In our case, they are the hydrogen ions  bearing positive charge. Observe that a positively charged ion in the solution is accompanied by a host of negatively charged ions. For that reason, the ion being accompanied by the negative ions (the ion dressed in a coat) has an increased mass. This fact is marked by a subscript asterisk.

 There are internal and external forces acting on the liquid.
  
 The internal forces, captured by curly bracket in Eq.~(\ref{eq=3}),
  arise because the liquid undergoes deformations at motion.
 The first internal force is represented by the pressure gradients, while the second is  dissipative, arising due to the viscosity flow.

The external force is the term ${\vec f}={\vec{\mathit F}}/\Delta V$, expressed as the  force per the unit volume, $\Delta V$. 
This force consists of two terms, ${\vec f} = {\vec f}_{1} + {\vec f}_{2}$.

 Since the fluid contains charged ions, the first force is the Lorentz force
\begin{equation}
  {\vec f}_{1} = \rho_{q}{\vec E} + {\vec J}_{q}\!\times\!{\vec B}.
\label{eq=6}
\end{equation}
 Here, $\rho_{q}=q/{\Delta V}=eN/{\Delta V}=e\rho$ is the charge $q=eN$ per the unit volume ($e$ is the electron charge), 
${\vec J}_{q} = q{\vec{\mathit v}}/{\Delta V}=e\rho{\vec{\mathit v}}$ is the density current, 
${\vec E}$ is an external electric field, and ${\vec B}$ is a magnetic field.

 The second force, we believe, is a conservative force ${\vec f}_{2} = -\nabla U_{2}({\vec{\mathit r}},t)$, 
 and we shall represent the potential $U_{2}$ as consisting of two potentials
 $U_{2,0}({\vec{\mathit r}},t)$ and $U_{2,1}({\vec{\mathit r}},t)$. 
  The first potential needs neutralization of the dissipative term $\mu\,\nabla^{\,2}{\vec {\mathit v}}$
  ($\mu$ is the dynamical viscosity coefficient).
  It is desirable that the difference between these two  terms, $U_{2,0}({\vec{\mathit r}},t)$ 
  and $\mu\,\nabla^{\,2}{\vec {\mathit v}}$,  would oscillate around zero (this will be considered further). 
  The second potential serves for searching and organizing the memory traces. 
Its contribution is to organize reallocating ion flows so that the memory traces surface in the brain. 
For those purposes, laws of the quantum mechanics could be useful.
  
  To apply the laws of quantum mechanics, we modified slightly the pressure gradient term.
  Let it be subjected  to the following modification
$\nabla P \rightarrow\nabla P - P \nabla\ln(\rho_{_{M}})= \rho_{_{M}}\nabla(P/\rho_{_{M}})$~\citep{Sbitnev2016b}.
 It will be important for us when we begin to derive the Schr\"odinger equation.
For that reason we need to derive the quantum potential, which is a main term underlying
  Bohmian quantum mechanics~\citep{Bohm1952a, Bohm1952b, Wyatt2005, BensenyEtAl2014}.
 
\subsection{\label{sec3A}Quantum potential}

  We begin with the modified pressure gradient
\begin{equation}
\label{eq=7}
  \rho_{_{M}}\nabla\biggl(
                   {{P}\over{\rho_{_{M}}}}
                   \biggr)
=   \rho\nabla\biggl(
                   {{P}\over{\rho}}
                   \biggr)
 = \nabla P - P\nabla\ln(\rho_{_{}}).
\end{equation} 
  The first term, the pressure gradient $\nabla P$, is represented in the usual Navier-Stokes equation~\cite{LandauLifshitz1987, KunduCohen2002}, 
 whereas, the second term, $P\nabla\ln(\rho_{_{}})$, is an extra term 
 describing change in the logarithm of the density distribution $\rho$ on the infinitesimal increment of length multiplied by $P$. 
 This means, that the change of pressure is due to the change in entropy per length, or else at negative sign the change in information content  per length because of flows of the sub-particles~\citep{Sbitnev2009}.

 First,  let us suppose that the pressure $P$  consists of the sum of two pressures $P_1$ and $P_2$. 
 
 As for pressure $P_1$ we begin with Fick's law
 which says that the diffusion flux, $\vec{\mathit J}$, is proportional to the negative value of the density gradient, 
 ${\vec{\mathit J}} =-D\nabla\rho_{_{M}}$~\citep{Grossing2010}.
 Since ${D}\nabla{\vec{\mathit J}}$ has dimension of the pressure, we define the first pressure:
\begin{equation}
\label{eq=8}
  P_1 = {D}\nabla{{\vec{\mathit J}}}
 = -D^{2}\nabla^{2}\rho_{_{M}}.
\end{equation}
 Here, $D$ is the diffusion coefficient.
 
 Observe that the kinetic energy of the diffusion flux~\citep{Grossing2010} of the liquid is 
 $(M/2)({\vec{\mathit J}}/\rho_{_{M}})^{2}$.
 It means that there exists one more pressure
 as the average momentum transfer per unit volume:
\begin{equation}
\label{eq=9}
  P_2 = {{M}\over{2\Delta V}}\biggl(
                                      {{{\vec{\mathit J}}}\over{\rho_{_{M}}}}
                               \biggr)^2
 = {{D^2}\over{2}}{{(\nabla\rho_{_{M}})^2}\over{\rho_{_{M}}}}.
\end{equation}
 Now we can see that, the sum of the two pressures, $P_1 + P_2$, divided by $\rho$ (we remark that $\rho_{_{M}} = m_{*}\rho$, see Eq.~(\ref{eq=5}))  reduces to the quantum potential~\citep{Sbitnev2016b}
\begin{equation}
\label{eq=10}
  Q =  {{P_{2}+P_{1}}\over{\rho}}
     =   m_{*} D^{2}
     \Biggl(
       {{1}\over{2}}\biggl(
                                   {{\nabla\rho}\over{\rho}} 
                                   \biggr)^2
        - {{\nabla^2\rho}\over{\rho}} 
     \Biggr).
\end{equation}

 Now we need to define the diffusion coefficient $D$ for the hydrogen ions in the physiological solution.
 This coefficient
% https://www.comsol.com/multiphysics/diffusion-coefficient 
  for the Brownian motion of a particle in the viscous liquid is described by the Stokes-Einstein formula~\citep{Miller1924}
\begin{equation}
\label{eq=11}
   D = {{k_{_{B}}T}\over{6\pi\mu r_{1}}}.
\end{equation}
 Here the denominator $6\pi\mu r_{1}$ is the Stokes  drag coefficient for the diffusing particles of radius $r_{1}$ in the fluid with the viscosity $\mu$.
 
 Since the particle has a charge, it drags a cloud of opposite charged particles. 
Therefore, there is an added drag coefficient, which is as follows:
\begin{equation}
   {{R}\over{r_{2}}}\cdot {{e^2}\over{2\pi\varepsilon_{w}r_{2}^2}}.
\label{eq=12}
\end{equation}
 Here $R$ is the linear resistance (its dimension is $\Omega\cdot$m), $e$ is the particle charge, $r_{2}$ is a cross-section size of the cloud, and $\varepsilon_{w}=81$ is the relative dielectric constant of sea water. Of great value of the dielectric constant are the peculiarities of the chemical structure of the H$_2$O molecule~\citep{Chaplin2016a}. 
It is connected to the fact that water is a strongly polar liquid and therefore has a soft orientational degree of freedom (rotation of molecular dipoles).  
Water possesses a large shielding effect. It means that two ions placed in the water lose electric sensibility towards each other through shorter distance than in vacuum.
    The cross-section of the shielding cloud is temperature-dependent.

   Following to the above observation, we define the diffusion coefficient as follows
\begin{equation}
 D = k_{_{B}}T\Biggl(
    {{1}\over{6\pi\mu r_{1}}} + 
    \biggr({{Re^2}\over{2\pi\varepsilon_{w}r_{2}^3}}\biggl)^{-1}  
                      \Biggr).
\label{eq=13}
\end{equation} 
 Observe that  in this equation inverse values of the drag coefficients are summed.\\

%\begin{widetext}
  TABLE~1: Some parameters of seawater with the salinity $S=10$ g/kg for  different temperatures
  (normal physiological saline  has the salinity   $S = 9$ g/liter)
\begin{center}
  \begin{tabular}{ l  |  c  c  c  c  r} % { l || c | c| c| c|| r}
    \hline
    \hline
  ~$T$, $^{\bf o}$K~~~~                                                &
  ~~~~$\mu^{\rm(a)}$, Pa$\cdot$s~~~~         &
    ~~~~$R^{\rm(b)}$, $\Omega\cdot$m~~~~ & 
     ~~$\varepsilon_{w}^{\rm(c)}$~~                &
     ~~~~$r_{1}^{\rm(d)}$, pm~~~~~             &
       $D^{\rm(e)}$, m$^2$/s~                          \\
    \hline
 273.15  &   $ 1.88\cdot10^{-3}$  &   $1.09$   &  80 & 53   &  $5.61\cdot10^{-9}$  \\
 298.15  &   $ 9.6\cdot10^{-4}$    &  $0.587$  &  80 & 53   &  $9.31\cdot10^{-9}$  \\
    \hline
    \hline
  \end{tabular}
  \vspace{4pt}
  {\large
  \begin{flushleft}
  (a)~\url{http://www.engineeringtoolbox.com/sea-water-properties-d_840.html}
  (b)~\url{http://www.comsistel.com/HAM%20documents/seawater.pdf},  ~$R=1/C$  %  Resistance = 1/Conductivity
  (c)~\url{http://www.civil.utah.edu/~cv5450/dielectric_const.html}
  (d)~\url{http://periodictable.com/Properties/A/AtomicRadius.v.html}
  (e)~\url{http://www.soest.hawaii.edu/oceanography/faculty/yhli/1974a.pdf}
  \end{flushleft}
  }
\hspace{24pt}  
\end{center}
%\end{widetext}

Given the parameters in Table~1, one can choose the cloud radius $r_{2}$, such that the diffusion coefficient $D$ would give the value shown in the same table.
In the case of $r_{2} = 691$ pm, we find $D = 5.61\cdot10^{-9}$ m$^2$/s for $T = 273$ K. 
For $r_{2} = 609.7$ pm we get $D = 9.31\cdot10^{-9}$ m$^2$/s for $T = 298$~K.
So the cloud radius $r_{2}$ represents a temperature-dependent parameter. The size of the cloud is decreased as the temperature rises. One can guess that the decreasing size of the cloud occurs because of thermal chaotization of the polar groups of the water molecules.

Now we need to define the diffusion coefficient $D$ in terms related to the quantum mechanics problems.
 These problems were considered by E. Nelson in his monographs~\citep{Nelson1967, Nelson1985}.
 According to~\citet{Nelson1966} Brownian motion of a sub-particle in ether is described by the Wiener process with the diffusion coefficient  equal to 
$\hbar/2 m$.

\subsection{\label{sec3B}Quantum diffusion coefficient}

In accordance with definition given in~\citep{Nelson1966},
 we write down the  diffusion coefficient as follows
\begin{equation}
  D = {{\bbar}\over{2m_{*}}}.
\label{eq=14}
\end{equation}
 Here $\bbar=b/(2\pi)$  is the reduced thermal action parameter~(\ref{eq=1}), like the reduced Planck constant $\hbar=h/(2\pi)$.
  And $m_{*}$ is an effective ion mass.
  We may evaluate the effective ion mass 
  by equating Eqs.~(\ref{eq=13}) and~(\ref{eq=14}). We obtain
\begin{equation}
  m_{*} = \frac {\delta\tau(2\pi)^{-1}}
   {2\Biggl(
    {\displaystyle {{1}\over{6\pi\mu r_{1}}} + 2\pi{{\varepsilon_{w}r_{2}^3}\over{Re^2}}}
                      \Biggr)}.
\label{eq=15}
\end{equation} 
  Here, ${\delta\tau}$ is $2\cdot10^{-13}$~s for the average lifetime of the hydronium ion H$_3$O$^+$~\citep{Bell1973}.
  From here we find that  $m_{*} \approx2.44\cdot10^{-27}$~kg for $T = 273$ K,    and 
   $m_{*}\approx1.83\cdot10^{-27}$~kg  for  $T=298$ K (room temperature).
  Comparison with the proton mass shows
  that the first effective mass is in about 1.5 times larger and
   the second is 1.1 times larger
   than the proton bare mass, $m\approx1.67\cdot10^{-27}$~kg.
 
    Compare the calculated values of the mass $m_{*}$ done at room temperature by the formulas~(\ref{eq=2}) and~(\ref{eq=15}). A relatively good coincidence of the numbers is striking.
  
\section{\label{sec4}Transition to the Schr\"{o}dinger equation}

Let us first rewrite the modified Navier-Stokes equation~(\ref{eq=3})  in the following presentation
\begin{widetext}
\begin{equation}
m_{*}\biggl(
             {{\partial {\vec {\mathit v}}}\over{\partial\,t}}   
             + \underbrace{{{1}\over{2}}\nabla {\mathit v}^{2} + [{\vec\omega}\times{\vec {\mathit v}}]}_{(a)}
         \biggr)
  + \underbrace{e  \biggl(
              {{\partial {\vec {\mathit A}}}\over{\partial\,t}}  + \nabla \phi
                           \biggr) + e  [{\vec B}\times{\vec {\mathit v}}]}_{(b)}
+\underbrace{\nabla \Bigl(
                                     Q  +  U_{2,1}
                               \Bigr)}_{(c)}
-\underbrace{\nabla \Bigr(
                                       m_{*}\nu \nabla  {\vec {\mathit v}} - U_{2,0}
                               \Bigl)}_{(d)}  = 0                               .
\label{eq=16}            
\end{equation}
\end{widetext}
In this equation instead of the mass density $\rho_{M} = m_{*}\rho$,
 the mass $m_{*}$ is represented as the main multiplier of the first brackets, 
 since we divided the equation on the density distribution~$\rho$.

The term captured by curly bracket (a) comes from $({\vec {\mathit v}}\cdot\nabla){\vec {\mathit v}}$.
The vector ${\vec\omega} = [\nabla\times{\vec{\mathit v}}]$ is called vorticity. The vorticity is perpendicular to the plane of rotation of the fluid~\citep{Lighthill1986}. 
It corresponds to the axis about which a vortex rotates~\citep{Sbitnev2015b}.

The term given to curly bracket~(b) is the {\it Lorentz force}~(\ref{eq=6}), 
but here we have expressed the electric field $\vec E$ through the vector, $\vec A$, and scalar, $\phi$, potentials:
${\vec E} = -\nabla\phi-\partial {\vec A}/\partial\,t$ (here, we adopted the SI unit).
Also, we take into account ${\vec J}_q\!\times\!\vec B = -\vec B\!\times\!{\vec J}_q$.

The gradient $\nabla(Q+U_{2,1})$ enclosed by curly bracket~(c) determines a generalized force of  Newton's second law~\citep{BensenyEtAl2014, Fiscaletti2012} in the Bohmian mechanics due to presence of the quantum potential $Q$.
This force is caused by both the quantum potential $Q({\vec r}, t)$
 and the imposed external potential $U_{2,1}({\vec r}, t)$.
\begin{figure}[htb!]
  \centering
  \begin{picture}(200,160)(30,10)
      \includegraphics[scale=0.4]{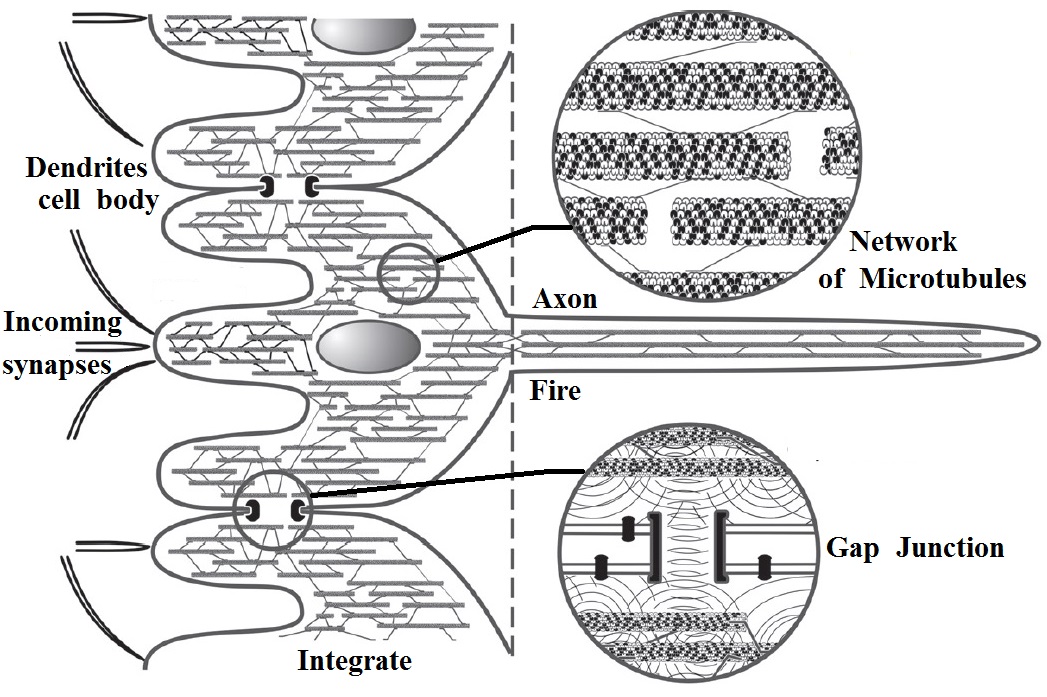}
  \end{picture}
  \caption{
  An 'integrate-and-fire' brain neuron and parts of other such
neurons are shown schematically with internal microtubules interconnected
by microtubule-associated proteins. The figure is taken from~\citep{Hameroff1998}.
  }
  \label{fig=5}
\end{figure}
The latter is due to abundant presence of the microtubules in neural cells, see Fig.~\ref{fig=5}, and other cell elements  in the brain tissues. 

 It is well known that neuronal filaments penetrate all corners of the brain so tightly that on thin brain slices a physiologist can sometimes not recognize a filamentous organization of the neuronal tissue. From here, it is important to emphasize that all functions of the brain are subordinated to service the vast water basin of the brain, regulation  of the water clusters, and the water 'wires' (proton pathways). 

The term captured by curly bracket~(d)  in Eq.~(\ref{eq=16})  contains energy dissipation 
$m_{*}\nu\nabla{\vec{\mathit v}}$ ($\nu = \mu/\rho_{M}$ is the kinetic viscosity coefficient). In order to neutralize this dissipation, the potential energetic term $U_{2,0}({\vec r},t)$ is added. Its role is to reset this dissipation to zero.  
This may be achieved by trembling of membranes and microtubules, 
the first due to regular working of ion pumps, the second to organization of transport of protein and food in the cells.
Perhaps, fluctuations about zero remain, so we may write:
\begin{equation}
  m_{*}{\tilde\nu}(t) \nabla^{2} {\vec{\mathit v}} = m_{*}{\nu} \nabla^{2} {\vec{\mathit v}} - \nabla U_{2,0}
\label{eq=17}
\end{equation}
 and $\langle {\tilde\nu}(t) \rangle = 0$, $\langle {\tilde\nu}(0){\tilde\nu}(t) \rangle > 0$.
  The water liquid acquires signs like superfluidity.
This phenomenon is represented as the result of the formation of a condensate consisting 
of a macroscopic number of hydrogen ions in a single quantum state that are subordinate to quantum laws.  

\subsection{\label{sec4A}Eightfold paths of the hydrogen ion and effect of superconductivity}

Theoretical aspects of superfluidity and superconductivity 
are well set out in~\citep{Vasiliev2013b, Vasiliev2015}.
A main idea is that the ordering of the zero-point fluctuations is a cause of the emergence of the Bose-Einstein condensate. 
 Let us again look on the Grotthuss mechanism in the context of transition to  the superconductivity. As seen in Fig.~\ref{fig=3}, once a hydrogen ion has passed in one direction, the other ion cannot pass. However, the latter can go along the same water-wire in the opposite direction, meaning that the Grotthuss mechanism switches the water-wire in backward and forward directions after each was passed by the hydrogen ion.

\begin{figure}[htb!]
  \centering
  \begin{picture}(200,130)(140,30)
      \includegraphics[scale=0.8]{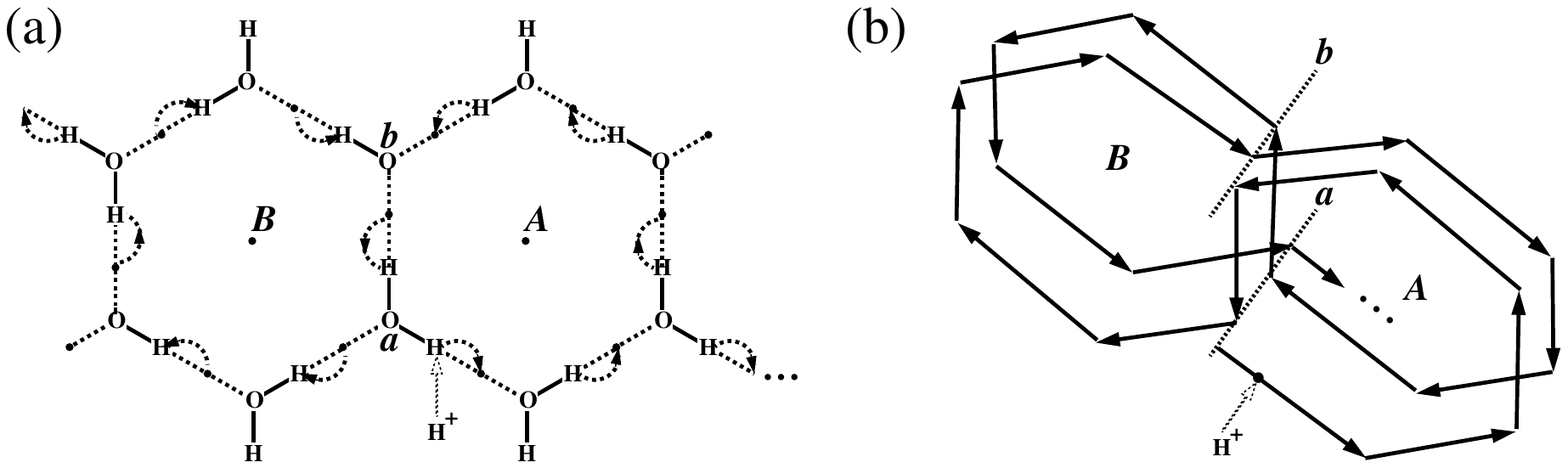}
  \end{picture}
  \caption{
 The eightfold path: 
 (a) according to the Grotthuss mechanism,  by hopping a hydrogen ion bypasses the hexagonal circuits around {\it A} and {\it B}  and changes the direction of the bypass on the bifurcation nodes {\it a} and {\it b};
 (b)  a conditional diagram showing the  journey of the hydrogen ion along the hexagonal circuits marked by {\it A} and {\it B}. 
The path bifurcates on the nodes {\it a} and {\it b} each time as soon as the hydrogen ion reaches them
(the dotted lines conventionally depict also a shift in time).
  }
  \label{fig=6}
\end{figure}

It turns out that there can be such an organization of water, when the Grotthuss mechanism is able to support a long-living hydrogen ion  current~\citep{HassanaliEtAl2013}. This organization is due to hexagonal circuits that belong to the fourth phase of water predicted by~\citet{Pollack2001, Pollack2013}.
Long-lived current can exist due to realization of the eightfold path of the hydrogen ion in the paired hexagonal circuits, see Fig.~\ref{fig=6}(a).
 In this organization all hexagonal circuits form a giant hexagonal packed circuit, where the currents circulate around the hexagonal circuits by the Grotthuss mechanism.
 The hydrogen ion travels on the water-wire passing through two hexagonal circuits, reproducing the eightfold path, Fig.~\ref{fig=6}(a), in such a way that  an open path for the ion always exists.
 
In order to clarify said above, we suggest taking a look on Fig.~\ref{fig=6}(b) which shows conditionally the  journey of the hydrogen ion along two hexagonal circuits marked by {\it A} and {\it B}.
 Here two bifurcation nodes, {\it a} and {\it b}, which also demonstrate conditionally a shift in time, are nodes where the hydrogen ion changes direction because of the closed other direction. One can clearly see in this diagram that the hydrogen ion can hop on this path so long as possible.

The hexagonal organization of water possesses by  a series of properties experimentally discovered by Pollack's team~\citep{ZhengEtAl2006, ChaiPollack2010, YooEtAl2011}. Because of these properties, the fourth state of water was named the exclusion-zone (EZ) water. Here we mention two particularly important properties for our work: (a)~EZ, which is more viscous for foreign molecules (moreover, it pushes away foreign molecules outside of the exclusion zone); and (b)~EZ, which has a negative charge. These properties provide favorable conditions for the Gutthuss mechanism. Observe first that the negative charge of EZ indicates the existence of many holes - empty seats with a negative charge where the hydrogen ion may hop. Hence it makes sense to consider the Gutthuss mechanism as counter-movements of the hydrogen ions and their holes,
as shown in Fig.~\ref{fig=4}. 
\begin{figure}[htb!]
  \centering
  \begin{picture}(200,80)(30,15)
      \includegraphics[scale=0.8]{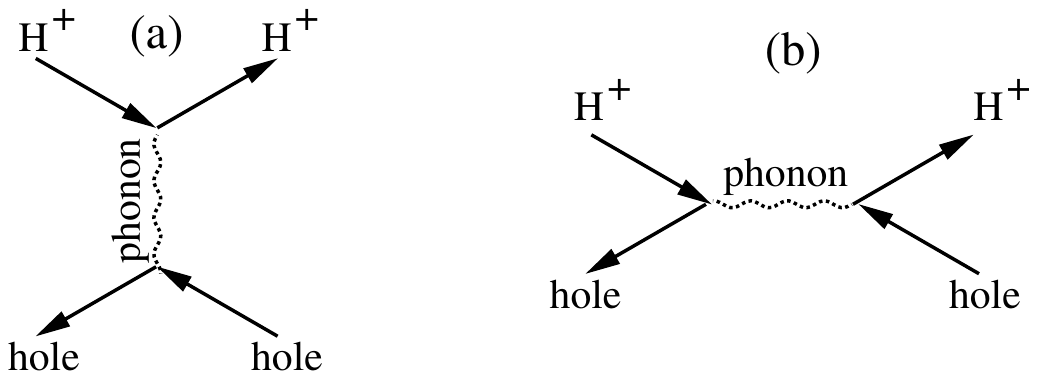}
  \end{picture}
  \caption{
 Two simple Feynman diagrams: (a) elastic H$^+$-hole scattering; (b) tunneling effect (hydrogen ion and hole annihilate producing a phonon. Next,  after some time, the phonon initiates emergence of pair of the hydrogen ion and hole).
  }
  \label{fig=7}
\end{figure}
\begin{figure}[htb!]
  \centering
  \begin{picture}(200,40)(-10,10)
      \includegraphics[scale=0.8]{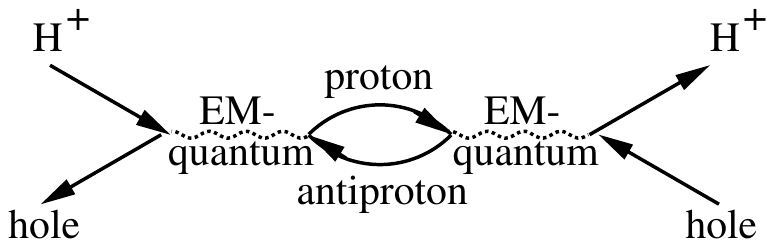}
  \end{picture}
  \caption{
  Feynman diagram showing migration of a hydrogen ion for long distances  through the superfluid quantum space.
  EM-quantum is an energy quantum of ElectroMagnetic radiation.
  }
  \label{fig=8}
\end{figure}
  The Feynman diagram technique~\citep{FeynmanHibbs1965, Jishi2013} suggests in this case that the hydrogen ions move from the past to the future, and the holes move in opposite direction - from the future to the past.
Fig.~\ref{fig=7} shows two simple Feynman diagrams, where phonon is a mediator between hydrogen ion and hole.
Figure (a) shows the elastic scattering of a hydrogen ion on a hole with the exchange by phonon. 
A classical counterpart is collision of two billiard balls.
Figure (b) shows the tunneling of a hydrogen ion through an obstacle. Initially, the hydrogen ion annihilates with a hole creating phonon. Then, the phonon, overcoming the obstacle, generates the hydrogen ion together with the hole.

Since the thermal action parameter is close to the Planck constant, we can envision an interesting Feynman diagram, shown in Fig.~\ref{fig=8}. Here, in the left section, the hydrogen ion annihilates with the hole, after which the process dives into the superfluid quantum space, where it may travel for a very long time. Next, this tunneling process again creates the hydrogen ion and the hole, shown in the right section, perhaps far from its dive into the superfluid quantum space. Such a tunneling process can provide a possible explanation for such subtle paranormal phenomena as, for example, Jung's synchronicity~\citep{Jung2008}, which manifests itself most conspicuously in the form of meaningful coincidences.
 
All paired hexagonal circuits (see Fig.~\ref{fig=6}) can be combined in a general~\citep{Pollack2001, Pollack2013} where the traveling of hydrogen ions along the aforementioned eightfold paths can exist for a very long time. Such an ensemble of paired circuits, supporting the Grotthuss mechanism, can represent the Bose-Einstein condensate, which exists at room temperature.  
As for the case of emergence of the Bose-Einstein condensate in EZ water, an important role, analogous to role of the zero-point fluctuations in superfluid  helium~\citep{Vasiliev2013b, Vasiliev2015}, can play  an infrasound induced by external sources in the physiologically significant frequency range of about 2 to 20~Hz.  When a wide space occupied by EZ water exists, the pairs of the hydrogen ions and the holes on this space can be assembled in a big coherent ensemble under effect of the  infrasonic waves at small decibels.
The infrasound sources can be not only of natural origin, but can also be arranged by people. For example, the transcendental meditation techniques and Vedic sounds have a significant effect on creating brain wave coherence~\citep{Nader2015}.

 Such bifurcation transitions from  unordered states to ordered show many common signs with those observed in experiments performed with small droplets bouncing on a fluid surface~\citep{CouderForte2006, Bush2015a}
When Faraday waves (with frequency about 80 Hz and with arbitrarily small amplitudes) are induced on this surface, the droplets begin to show coherent behavior through formation of a synergistic complex: droplet \& wave = walker~\citep{EddiEtAll2011, Bush2015a}. 
The walkers behave as quantum particles  guided by waves, by demonstrating interference effects at passing through gates in screens~\cite{CouderForte2006}.

\subsection{\label{sec4B}Irrotational and solenoidal vector fields}

  Let us continue the mathematical transformations.
   The fundamental theorem of the vector calculus, Helmholtz's theorem, states that any vector field can be expressed through   the sum of irrotational and solenoidal fields.
The current velocity ${\vec{\mathit v}}$ can be represented as consisting of two components --  irrotational and solenoidal~\cite{KunduCohen2002}:
\begin{equation}
\label{eq=18}
  {\vec{\mathit v}} = {\vec{\mathit v}}_{_{S}} + {\vec{\mathit v}}_{_{R}},
\end{equation}
 where subscripts $S$ and $R$ point to the existence of scalar 
  and vector (rotational) potentials  underlying the emergence of these  velocities~\citep{Sbitnev2016b}.
 These velocities relate to vortex-free and vortex motions of the fluid medium, respectively. 
 They satisfy the following equations
\begin{equation}
\label{eq=19}
 \left\{
    \matrix{
           (\nabla\cdot{\vec{\mathit v}}_{_{S}}) \ne 0, & [\nabla\times{\vec{\mathit v}}_{_{S}}]=0, \cr
           (\nabla\cdot{\vec{\mathit v}}_{_{R}})  =  0, &\, [\nabla\times{\vec{\mathit v}}_{_{R}}]={\vec\omega}. \cr
           }
 \right.
\end{equation}

One can trace parallels between these equations and the Maxwell equations for electric and magnetic fields~\citep{Martins2012, MartinsPinheiro2009}. 
In this key, we may represent the kinetic momentum 
${\vec p}=m_{*}{\vec{\mathit v}}$ and the kinetic energy by the following expressions
\begin{equation}
\label{eq=20}
 \left\{\,
    \matrix{
           {\vec{\mathit p}} = m_{*}({\vec{\mathit v}}_{_{S}} + {\vec{\mathit v}}_{_{R}})
            = \nabla S - e{\vec A},
\cr\cr
           {\displaystyle
            m_{*}{{({\mathit v}_{_{S}} + {\mathit v}_{_{R}})^2}\over{2}}} = 
           {\displaystyle{{1}\over{2m_{*}}}}(
                       \nabla S - e{\vec A}
                                        )^2 .
\cr
           }
 \right.
\end{equation}
 Here, $S$ is a scalar function called the {\it action}, 
 the gradient of which determines the irrotational velocity ${\vec{\mathit v}}_{_{S}}$, 
whereas the vector potential underlies determination of the solenoidal velocity ${\vec{\mathit v}}_{_{R}}$. 

 Taking into account $m_{*}{\vec{\mathit v}} =\nabla S - e{\vec A}$ let us rewrite Eq.~(\ref{eq=16})  in detail
\begin{eqnarray}
\nonumber 
&&
   {\frac{\partial\;}{\partial\,t}} (\nabla S - e{\vec A}) 
+ {\frac{1}{2m_{*}}}\nabla (\nabla S - e{\vec A})^2
\\
&+&
   e  {\frac{\partial\;}{\partial\,t}} {\vec A} + e\nabla\phi
 + m_{*}[{\vec\omega}\!\times\!{\vec{\mathit v}}]
 + e[{\vec B}\!\times\!{\vec{\mathit v}}]
- m_{*}{\tilde\nu}(t)\nabla^{2}{\vec{\mathit v}}
+ \nabla(Q + U_{2,1}) = 0.
\label{eq=21}
\end{eqnarray} 
  First, one can see that the both terms $e{\partial {\vec A}}/{\partial t}$ in this expression kill each other.
 Also, the term  $m_{*}[{\vec\omega}\!\times\!{\vec{\mathit v}}]$ can be transformed to 
$-e[{\vec B}\!\times\!{\vec{\mathit v}}]$ 
 as soon as  instead of ${\vec\omega}$  we substitute
  its representation $[\nabla\!\times\!{\vec{\mathit v}}_{_{R}}]= - e/m_{*}[\nabla\!\times\!{\vec A}]$
  following from~(\ref{eq=20}): ${\vec{\mathit v}}_{R}=-e/m_{*}{\vec A}$.
 From here we see that contribution of the magnetic field in a cognitive process of the brain disappears:
  $e[{\vec B}\!\times\!{\vec{\mathit v}}]-e[{\vec B}\!\times\!{\vec{\mathit v}}]=0$. 
  
 It is interesting to note the following unusual phenomenon stemming from this equation. Suppose that the vorticity ${\vec\omega}$ is not compensated by the magnetic field. Then, this equation can also 
 give vortex solutions~\citep{Sbitnev2015b, Sbitnev2016c}. 
 One can imagine a helicoidal vortex ring rising over the head that resembles the torus.  
The hydrogen ions rotating about the torus radius burn down in the atmosphere. 
As a result, it will look as a luminous cloud over the head, called halo. 
  
\subsection{\label{sec4C}Final transformations}

Now we can rewrite Eq.~(\ref{eq=21}) in a compact form
\begin{equation}
 \nabla
 \biggr\{
      {\frac{\partial\;}{\partial\,t}} S + {\frac{1}{2m_{*}}} (\nabla S - e{\vec A})^2 + e\phi
     + {\tilde\nu}(t) m_{*} f(\rho)
     + Q  + U_{2,1}
 \biggl\} = 0.
\label{eq=22}
\end{equation}
The term $f(\rho)=\nabla{\vec{\mathit v}}=-d\ln(\rho)/d\,t$~\citep{Sbitnev2016b}
comes from the continuity equation of the mass density Eq.~(\ref{eq=4}),
 which is rewritten as the continuity equation for the density distribution
\begin{equation}
 {{\partial\,\rho_{_{}}}\over{\partial\,t}} +(\nabla\cdot{\vec{\mathit v}})\rho_{_{}} = 0.
\label{eq=23}
\end{equation}
The expression enclosed in curly brackets in Eq.~(\ref{eq=22}) is the Hamilton-Jacobi-like equation
\begin{equation}
     {\frac{\partial\;}{\partial\,t}} S + {\frac{1}{2m_{*}}} (\nabla S - e{\vec A})^2 + e\phi 
     + {\tilde\nu}(t) m_{*} f(\rho)
     + Q  + U_{2,1}({\vec r})
  = C.
\label{eq=24}
\end{equation}\\
 loaded by the extra terms $Q$ and $ {\tilde\nu}(t) m_{*} f(\rho)$. 

            Eqs.~(\ref{eq=23}) and~(\ref{eq=24}) stem from the following nonlinear Schr\"{o}dinger equation
\begin{widetext}
\begin{equation}
  {\bf i}\bbar{\frac{\partial \Psi}{\partial\,t}} =
  {\frac{1}{2m_{*}}}\Bigr(
                                      -{\bf i}\bbar \nabla  - e{\vec A}
                               \Bigl)^2\Psi 
  +  
  \underbrace{ 
      e\phi\Psi
  +  U_{2,1}({\vec r})\Psi      
  +   {\tilde\nu}(t)m_{*}f(\rho)\Psi
  -   C\Psi  }.
\label{eq=25}
\end{equation}\\
\end{widetext}   
 It should be noted that in this equation the reduced thermal action parameter, $\bbar$, is presented,
  not the reduced Planck constant $\hbar$. 
In the above equations $C$ is the integration constant that arises at extracting solution~(\ref{eq=24}) from Eq.~(\ref{eq=22}).

We arrived to a single equation, the Schr\"{o}dinger-like equation, describing flows of the hydrogen ions in the water by their representations through the complex-valued function - the wave function $\Psi({\vec r},t)$. Observe that $\rho({\vec r},t)$ deals with the probability of detection of the hydrogen ion in the vicinity of point ${\vec r}$ in the moment of time $t$, and the action $S({\vec r},t)$ points to its mobility in the vicinity of this point.
 Substituting in the equation~(\ref{eq=25}) the wave function represented in the polar form
\begin{equation}
   \Psi = \sqrt{\rho}\exp\{{\bf i}S/\bbar\}
\label{eq=26}
\end{equation}
 and after separating the real and imaginary parts  of this equation,
  we come to the equations~(\ref{eq=24}) and~(\ref{eq=23}), respectively.

It is instructive to draw attention to the  Schr\"{o}dinger equation written phenomenologically by~\citet{BeckEccles1992} in 1992.
This equation, written for room temperature, can be true if only the superfluid medium is supported  (the viscosity is  absent). The time of the metastable instability of electronic transition, $\tau$, was evaluated by the authors about $10^{-13} - 10^{-14}$~s. 
Their estimation shows a relatively good agreement with the average lifetime of hydrogen ion, ${\delta\tau}=2\cdot10^{-13}$~s. 
In 2008~\citet{Beck2008} replaced hydrogen atom by electron, however.
This replacement was not a good idea. Electron is a light particle, and it is not robust with respect to thermal fluctuations  in a wet and warm medium. Therefore its transport in biological tissues occurs by capturing it by heavy molecules~\citep{LodishEtAl2000}. 
In the electron transport the following molecular complexes are involved - 
NADH dehydrogenase, cytochrome bc$_1$, cytochrome oxidase, and ATP synthase, synthesizing ATP. 

Now let us clarify some details of this equation in light of the cognitive activity of the brain.
 
\subsection{\label{sec4D}Features of the Schr\"{o}dinger-like equation}

The last four terms  covered by curly bracket  in Eq.~(\ref{eq=25}) describe a connection with the environmental medium. 

The first term, $e\phi({\vec r},t)$, describes  contribution of the electric potential $\phi({\vec r},t)$ in a state of neuron activity.  Voltage-gated proton channels are ion channels that open with depolarization in a strongly pH-sensitive manner~\citep{Decoursey2003}. 
These channels open only when the electrochemical gradient is directed outward. 
 The resting membrane potential of neural cells is about $-70$ mV. 
\citet{HodgkinHuxley1952} have given a strict mathematical description of the electric current across the membrane through the analysis of ion flows.
All this means, that the electric field is a main player in the cognitive brain functions,
whereas the magnetic field, as was shown above, is not critical to these functions.

The second term, $U_{2,1}({\vec r})$, is determined by architecture of the nervous tissue which can be complex enough. Complexity of the architecture of the brain is due to fractality of the neuron filaments penetrating any crannies of the brain~\citep{GardinerEtAl2010}. By itself, fractality of the central nervous system represents an amazing phenomenon of self-organized criticality~\citep{Werner2010} leading to organization of memory traces. Right up to organization of a set of microtubules, Fig.~\ref{fig=5}, and tubulin subunit dipole states~\citep{HameroffPenrose2014a}.

The third term, ${\tilde\nu}(t)m_{*}f(\rho)$, as stated above concerning Eq.~(\ref{eq=17}), is fluctuating about zero. 
In some way, this term introduces noise. So the  Schr\"{o}dinger-like equation~(\ref{eq=25}), 
which is analogous the Gross-Pitaevskii equation~\citep{Sbitnev2016b}  because of this term
 can be considered as the Langevin equation loaded by a source of the color noise. 
This term adds decoherence in a flow of the hydrogen ions. 

The fourth term, the integration constant $C$,
 can have arbitrary value by returning an uncertain phase shift of the wave function. 

\section{\label{sec5}The path integral: passage through the gap junctions}

\begin{figure}[htb!]
  \centering
  \begin{picture}(180,150)(20,10)
      \includegraphics[scale=0.7]{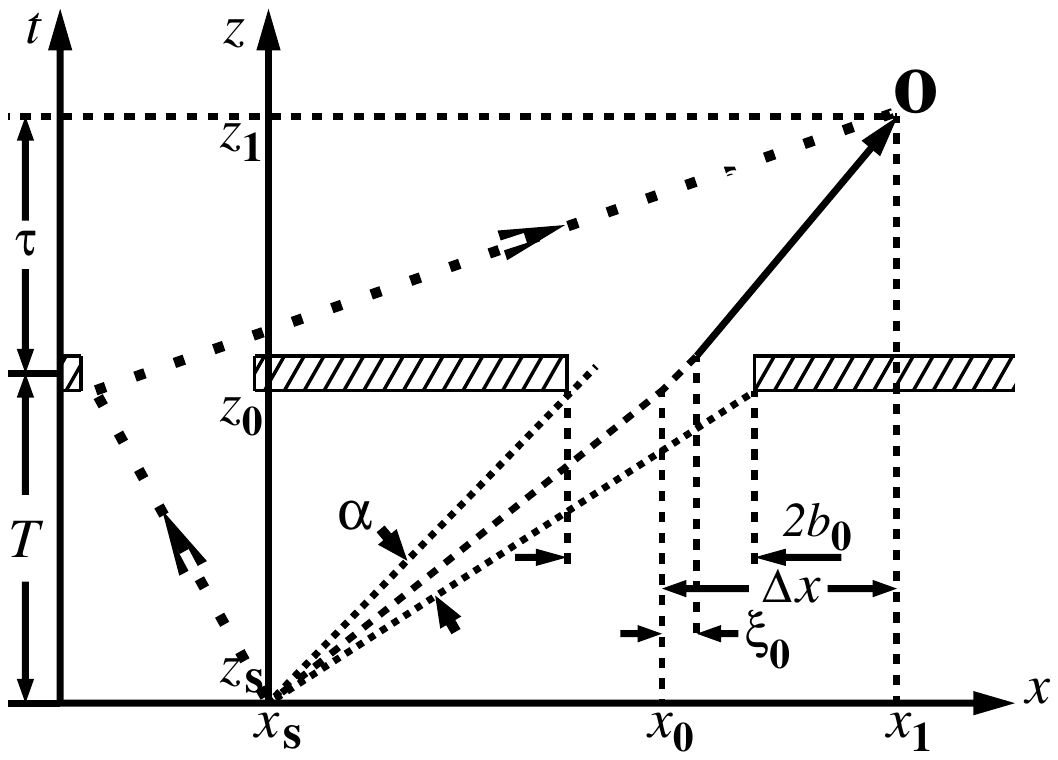}
  \end{picture}
  \caption{
 The path integral technique calculates all paths leading from a source located in point $(x_s,z_s)$
 to a point of observation marked by letter {\bf O}.
 It is shown by two slits, gap junctions, placed on a membrane surface.
  }
  \label{fig=9}
\end{figure}

There are rigorous mathematical derivations
concerning extraction of the Schr\"odinger equation out of the
 Feynman path integral~\citep{FeynmanHibbs1965, Derbes1996, Sbitnev2012}.
Following this statement, the Schr\"{o}dinger wave equation can be resolved by heuristic writing of a solution by using the Huygens' principle~\citep{Sbitnev2012},
 which mathematically appears as
\begin{equation}
  |\Psi({\vec r},t)\rangle= \int  K({\vec r},{\vec\xi};t)\Psi({\vec\xi},0)d{\vec\xi}.
\label{eq=27}
\end{equation}
 The propagator $ K({\vec r},{\vec\xi};t)$  bears information about the neuron tissue that is contained in the terms covered by curly brackets in Schr\"{o}dinger equation. 
Here the integral summarizes all paths leading from a source of radiation to a point of observation. Let it be conditionally signed by letter {\bf O}, Fig.~\ref{fig=9}.
Two slits, in our case gap junctions, shown in this figure give possibility to the hydrogen ions to pass to point {\bf O} via two pathways. Since these travels are described by wave function $|\Psi\rangle$, its contribution is evaluated by superposition of all pathways leading to the point {\bf O}.  An interference effect in this point can be either constructive or destructive. It means, that the hydrogen ion can either arrive into the point {\bf O} with the probabilty $\langle\Psi |\Psi\rangle$ or not.

The gap junctions are specialized intercellular connections between different cells directly connecting the cytoplasm of two cells, see the insert in Fig.~\ref{fig=5}.
\begin{figure}[htb!]
  \centering
  \begin{picture}(180,140)(33,0)
      \includegraphics[scale=0.55]{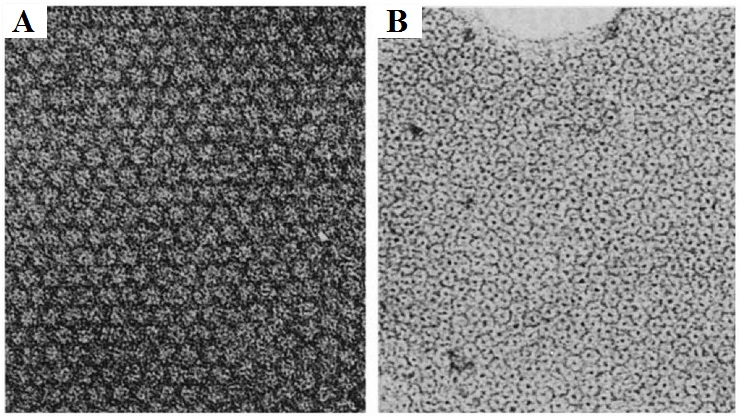}
  \end{picture}
  \caption{
   Gap junctions stained with
  (A) phosphotungstic acid, $\times$580,000;
  (B) and uranyl acetate, $\times$470,000.
  Well visible is the hexagonal packing.
 Photos are borrowed from~\citep{Zampighi1987}.
  }
  \label{fig=10}
\end{figure}
 The gap junctions, electrical synapses, exist in every major region of 
 the central nervous system~\citep{ConnersLong2004, SohlEtAl2005, MeierDermietzel2006}.
 The gap junctions are observed predominantly on the glial cells which are much larger than the number of neurons in the brain. They may contribute to the cognitive processes  as perception and attention~\citep{NagyEtAl2004}.

Gap junctional  intracellular communications  are formed into ordered arrays showing predominantly hexagonal packing, Fig.~\ref{fig=10}, with about 6 to 9 nm center-to-center spacing~\citep{Zampighi1987, BergEtAl2002}.
 Such  ordered arrays can be perfect slit gratings for the ion beams passing through them and reproducing an interference effect behind them.

 For the sake of simplicity we shall consider an ordered localization of the gap junctions 
along some membrane direction, say along the $x$-axis.
 And the coherent hydrogen ion beam is incident on the gap grating along the $z$ direction.
 
 Let the hydrogen ion beam be at the room temperature. $T=928$~K.
The de Broglie wavelength of the hydrogen ions, $\lambda_{dB} = b/({c_s}m_{*})$, is about 0.3 nm. 
Here  $b$ is the thermal action parameter, and at the room temperature, the sound speed $c_s = 1508$ m/s, and the ion mass
 $m_{*}\approx 1.83\times10^{-27}$~kg. 
Note that here we  are dealing with a sound wave in this water medium.

By applying the path integral technique for describing the wave propagation through a grating consisting of $N$ slits,
 we get the solution~\cite{Sbitnev2010b, Sbitnev2012}
\begin{equation}
% see d:\Valery\memristor\Tubulin\32slits2.xmcd
  |\Psi(x,z)\rangle =
 {{1}\over{\sqrt{\displaystyle 1 + {\bf i} {{z\lambda_{dB}}\over{2\pi w^{2}}} }}} 
 \cdot
   \sum\limits_{n=0}^{N-1}
  \exp
  \left\{
   \matrix{
   - {{\displaystyle \Biggl(x - \Biggl(n - {{N-1}\over{2}}
                                \Biggr)d
                     \Biggr)^{2}}\over{\displaystyle 2w^2\Biggr(
                                                                 1 + {\bf i} {{z\lambda_{dB}}\over{2\pi w^{2}}}
                                                         \Biggl)}}
          }
  \right\}.
\label{eq=28}
\end{equation}
 Here, $w$ is a width of the slit, $d$ is the distance between  slits, and $n$ is the sequence number of the slit. 
 The slits are placed along the $x$-axis, with equidistant spacing between them, 
  and the $z$ axis is perpendicular to the grating plane.

\begin{figure}[htb!]
  \centering
  \begin{picture}(180,170)(40,15)
      \includegraphics[scale=0.5]{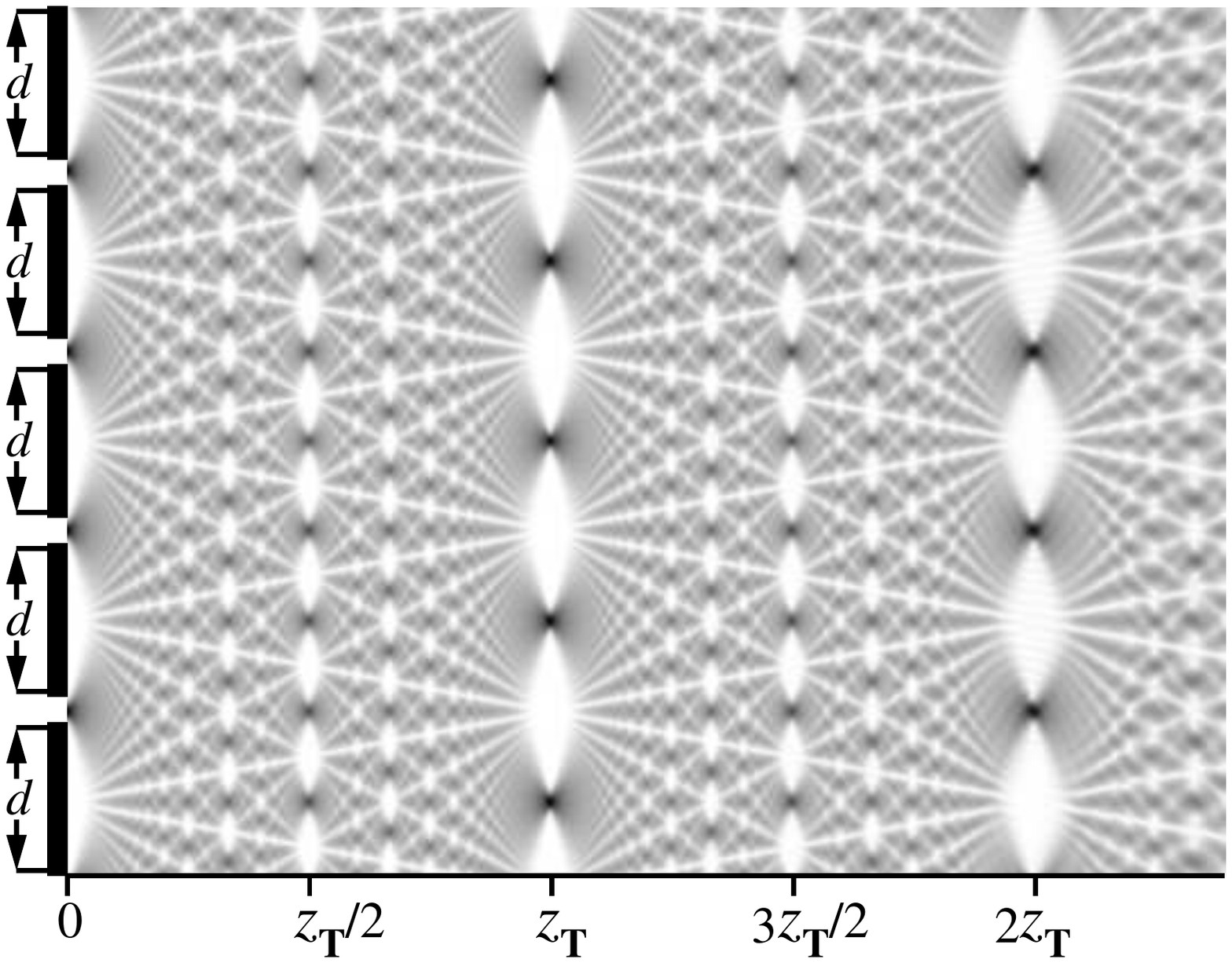}
  \end{picture}
  \caption{
 Interference pattern from $N=32$ slits, coherent beam,  ${\lambda_{dB}}=w = 0.3$ nm, $d=9$ nm, $z_{\rm T}= 270$ nm. 
 Cut only 4 central slits  in the near field shows the Talbot-like carpet.
  }
  \label{fig=11}
\end{figure}

The spacing between the slits let be 9 nm.
The result of the simulation, namely, the density distribution function $p(x,z)=\langle\Psi(x,z)|\Psi(x,z)\rangle$, 
is shown in Fig.~\ref{fig=11}. 
The dark gray regions in this figure show a high probability of finding the hydrogen ion in the vicinity of point $(x,z)$, whereas the light gray regions are those where the hydrogen ions are rare.

The characteristic spacing~\citep{BerryKlein1996}
\begin{equation}
   z_{\rm T} ={{d^{\,2}}\over{\lambda_{dB}}}
\label{eq=29}
\end{equation}
is a convenient length adopted in the interferometry at studying the interference patterns.
It is called the Talbot distance.
 In our case, it is about 270 nm.
So, for example, on the distance $z_{\rm T}$ the first self-image of the grating,  the most striking, appears shifted on the half of the grating period $d$ aside. 
On the distance $2z_{\rm T}$ a second self-image appears that is identical to positions of the slits in the grating, and so forth. 
The Talbot carpet, representing fractal, appears, when $N$ and $d$ tend to infinity~\citep{BerryKlein1996}.
In our case, however, $N$ and $d$ are finite numbers,  and the ion beam may not be coherent.
It means that the interference pattern will be blurred. Nevertheless, the first self-image can be reproduced with the some visibility~\citep{Sbitnev2013a}. 

It also means, if on a distance equal to $z_{\rm T}$, membrane surface of another neuron with relevant positions of the gap junctions exists, then the tunneling through such gap junctions~\citep{Hameroff1998} 
with amplification effects  can occur. 
Observe that inside the neurons a huge set of microtubules is located~\citep{CraddockEtAl2012}. 
So, each microtubule-associated protein  perturbs the coherent ion beam. These perturbations modify the wave ion field depending on what patterns from the heavy ions (predominantly the calcium ions) on the microtubules have been formed. 
It is a manifestation of the memristive effect~\citep{ Chua2011}.

\section{\label{sec6}Conclusion}

\citet*{StappEtAl2005} 
write:
"A principal function of the brain is to receive clues from the environment, to form an appropriate plan of action and to direct and monitor the activities of the brain and body specified by the selected plan of action".
Summarizing the above, we can say that 
consciousness is a form of motion allowing us to define an optimal path to achieve a goal. 

In classical mechanics, the optimal path for a ball moving along some complex relief follows from the least action principle~\citep{Lanczos1970}.
 Namely, the optimal path, along which the integral 
\begin{equation}
    S  = \int\limits_{0}^{t} L({\vec q}, {\dot{\vec q}},\tau) d\tau
\label{eq=30}
\end{equation}
is constant and where $ L({\vec q}, {\dot{\vec q}},\tau)$ is the Lagrange function.
But the ball does not know the foundations of the theoretical mechanics. 
 It always moves through the complex relief along the most optimal path. 
 How can the ball (an inorganic material object) find that path?
 
 Recall that at a place of contact of the ball with a complex relief surface there is a cohesion of these contacting surfaces. These contacting surfaces are not smooth but on the nano scale one can find enormous amounts of irregular protrusions and dents. 
They are sources of a noise perturbing the motion of the ball on small scales. 
The  small random perturbations initiate the Markovian process~\citep{Nelson1967},  
and as a result the  the ball  moves along the optimal path~\citep{Sbitnev2008}.

As for the quantum realm, Paul Dirac drew attention in 1933~\citep{Dirac1933}
 to a special role of the action $S$, which can exhibit itself in expressions through a term $\exp\{{\bf i}S/\hbar\}$. In particular, he had written in ~\citep{Dirac1945}:
  "we can use the formal probability to set up a quantum picture rather close to the classical picture in which the coordinates ${\vec q}$ of a dynamical system have definite values at any time. We take a number of times $t_1, t_2, t_3, \cdots$ following closely one after another and set up the formal probability for the $q$'s at each of these times lying within specified small ranges, this being permissible since the $q$'s at any time all commutate. We then get a formal probability for the trajectory of the system in quantum mechanics lying within certain limits. This enables us to speak of some trajectories being improbable and others being likely."  

Richard Feynman has formulated acceptable language for the description of moving quantum objects, where the decisive role has the term $\exp\{{\bf i}L{\delta t}/\hbar\}$~\citep{Feynman1948}. 
The idea is that the above term proposes mapping a wave function from one state to another through a small time interval $dt$. Feynman's genius insight has resulted in the understanding that the integral kernel (propagator) of the time-evolution operator can be expressed as a sum over all possible paths (not just the classical one), connecting the points ${\vec q}_a$ and ${\vec q}_b$ with the weight factor $\exp\{{\bf i}S({\vec q}_a; {\vec q}_b; t)/\hbar\}$:
\begin{equation}
  K({\vec q}_a; {\vec q}_b) = \sum\limits_{\forall~{\rm paths}}
   A \exp\{{\bf i}S({\vec q}_a; {\vec q}_b; T)/\hbar\}
\label{eq=31}
\end{equation}
 where $A$ is an normalization constant.
 So, all paths going from ${\vec q}_a$ to ${\vec q}_b$ are subjected to trials. A resulting contribution is given by the interference of all wave functions $ \exp\{{\bf i}S({\vec q}_a; {\vec q}_b; T)/\hbar\}$ in the final point ${\vec q}_b$.
  One may imagine that each point belonging to this physical scene is a source of a wavy radiation, and all these waves make a contribution to the interference at the final point ${\vec q}_b$. The interference may be either constructive or destructive. 
  In the first case, there is a high probability that the particle arrives at point ${\vec q}_b$. 
In the second case, the probability is almost zero and the particle does not arrive at this point.
 
Factually, the above picture agrees with the de Broglie idea about existence of the pilot wave guiding the particle from the initial point ${\vec q}_a$ to the final one ${\vec q}_b$~\citep{deBroglie1987} - the particle \& the pilot wave behave as one entity.
  The interference pattern in this case is defined from all subjects situated in the space from which reflected waves are returned to the particle.
 
Through the prism of the above description, we may now consider the consciousness evolving in the wet, warm, and noisy brain system. It interacts with a massive volume of memory stored in a deeper, finer-grained scale of a memristive system~\citep{Chus1971, Chua2011}. The interaction manifests itself through the destructive and constructive interference effects, like the effect of a vote. The memristive system by itself is based on the microtubules involving the calcium ions (networks of the granulated units of Ca$^{2+}$), including the calcium-calmodulin-dependent protein kinase II (CaMKII), which is implicated in the strengthening of active neural connections~\citep{CraddockEtAl2012}.

\begin{acknowledgements}
The author thanks Dr. Anatol Bragin and Dale Booth for useful and valuable remarks and suggestions.
The author thanks also Adam Crowl for an interesting remark.
The author is grateful to the reviewers for the constructive critique.
The author thanks Miss Pipa (Quantum Portal administrator) for writing a program that has calculated and drew Fig.~\ref{fig=11}.
\end{acknowledgements}

%===================================================
 
%\bibliography{D:/bibtex}

\end{document}